\documentclass[12pt]{article}
\usepackage{amssymb,amsmath,epsfig}
\allowdisplaybreaks

\begin{document}

\title{\bf Circular Motion and Energy Extraction in a Rotating Black Hole}

\author{M. Shahzadi$^1$ \thanks{misbahshahzadi51@gmail.com}, Z. Yousaf$^1$ \thanks{zeeshan.math@pu.edu.pk} and
Saeed Ullah Khan $^{2}$ \thanks{saeedkhan.u@gmail.com}\\
$^1$ Department of Mathematics, University of the Punjab,\\
Quaid-i-Azam Campus, Lahore-54590, Pakistan.\\
$^2$ Department of Mathematics, COMSATS University Islamabad,\\
Islamabad-45550, Pakistan.}
\date{}
\maketitle

\begin{abstract}

This paper explores the circular geodesics of neutral test particles
on an equatorial plane around a rotating black hole. After using
equations of motion of scalar-tensor-vector gravity with the
circular geodesics of null-like particles, we find the equation of
photon orbit. With the help of an effective potential form, we have
examined the stable regimes of photons orbits. The Lyapunov exponent,
as well as the effective force acting on photons, is also
investigated. We examine the energy extraction from a black hole via
Penrose process. Furthermore, we discuss the negative energy state
and the efficiency of energy extraction. We have made compare our
results with that obtained for some well known black holes models.
We concluded that the efficiency of the energy extraction decreases with
the increase of dimensionless parameter of theory and increases as
spin parameter increases.\\

\end{abstract}

\section{Introduction}

The process of exploring the hidden aspects of our Cosmos has always
been a source of great attractions for many researchers. Different
Cosmic data-sets indicate that our Universe is undergoing
accelerated expansion phase that has opened up new directions. It is
believed that this expansion is due to an enigmatic force, dubbed as
dark energy (DE). Apart from that, the dynamics of another
mysterious matter need to explore which is widely known as dark
matter (DM). Such a matter has no interaction with the light and
electromagnetic force. Modified gravity theories are useful to
reveal the incomprehensible nature of DE as well as DM. These
theories are the modification of the usual Einstein gravity (for
further reviews on DE and modified gravity, see, for
instance,~\cite{R3,R4,R9,R10}). Moffat \cite{1} proposed
scalar-tensor-vector gravity (STVG) as one of the models that could
be considered as an alternative for DM. In the action of this
modified theory of gravity (MOG), field for three scalars and
massive vector are included along with the Einstein-Hilbert term and
the matter action. This theory is useful to discuss the rotational
curves of galaxies, solar system, gravitational lensing of galaxy,
motion of galaxy clusters without considering DM \cite{2}.

Moffat and Rahvar \cite{3} studied some dynamical features of galaxy
clusters in the gravity by applying weak-filed approximations.
Moffat \cite{4} studied the shadows of black holes (BHs) in the same
theory and found that size of the shadows increases as the parameter
$\alpha$ increases. Mureika et al. \cite{5} discussed the
thermodynamics of MOG-BHs and found that the entropy area law
changes as $\alpha$ increases. Lee and Han \cite{6} investigated
that radius of the inner most stable circular orbit for Kerr-MOG BH
increases with the increasing values of $\alpha$. Sharif and
Shahzadi \cite{7} analyzed the effects of magnetic field on the
particle dynamics for timelike geodesics around Kerr-MOG BH. A useful
result on the the investigation of null geodesics in the background of MOG is produced by
Rahvar and Moffat \cite{7a}. They analyzed the propagation of electromagnetic
waves in MOG and calculated the corresponding deflection angle. They found quite larger
bending angle of light in MOG as compared to that obtained in general relativity.

In recent years, the motion of particles (massless or massive) near
a BH remained a compelling issue in BH astrophysics. The study of
geodesics reveals the geometrical structure as well as the important
features of a curved spacetime. They also help to determine the
gravitational field around a BH. Among various types of geodesics,
the most fascinating one is circular geodesics. The circular
geodesic motion of particles for Schwarzschild,
Reissner-Nordstr\"{o}m as well as Kerr BHs has been discussed
\cite{8}. Bardeen \cite{9} studied the properties of a Kerr BH and
its circular orbits. Pugliese and Quevedo \cite{10} identified the
regions inside the ergoregion of Kerr BH and deduce that
distribution of circular orbits depends on the rotation parameter of
a source.

A physical particle follows either timelike or null geodesics. The
study of circular null orbits is important from both theoretical and
astrophysical points of view. Frolov and Stojkovic \cite{11}
analyzed the particle as well as light motion and concluded that
there are no stable circular orbits in an equatorial plane around a
five-dimensional rotating BH. Konoplya \cite{12} studied the motion
of both massive and massless particles near magnetized BHs and found
that tidal force has strong effect on particle motion. Fernando
\cite{13} investigated that the circular orbits of photons around
Schwarzschild BH are more stable than for a BH surrounded by
quintessence matter. Khoo and Ong \cite{14} investigated that there
cannot be any photon orbit on the event horizon of a non-extremal
Kerr-Newman BH. Pradhan \cite{15} explored null circular geodesics
near Ay\'{o}n-Beato-Garc\'{i}a BH and found that at a certain radius, there
exists zero angular momentum orbits. Yousaf and Bhatti \cite{15a}
studied the stable regions of some relativistic compact structures in MOG.

Stuchl\'{i}k et al. \cite{15b} studied the particle collision and
circular geodesics in the braneworld mining Kerr-Newman naked
singularity spacetimes and found that the radius of stable circular
photon orbit is almost independent of the spin parameter $a$, being
situated nearly to the limiting radius $r=b$, (where $b$ is the
tidal charge parameter) and this radius corresponds to the radius of
innermost stable circular orbit. They also concluded that there
exist an infinitely deep gravitational well centered at the stable
photon circular orbit.

Energy extraction from a rotating BH is one of the most important
and interesting issues in general relativity as well as in
astrophysics. Penrose \cite{16} demonstrated a highly efficient
mechanism to extract the energy from a rotating BH and related to
the existence of negative energy in the ergoregion. Nozawa and Maeda
\cite{17} deduced that more energy can be extracted from higher
dimensional BHs as compared to the energy extortion from $4D$ Kerr
BH. Pradhan \cite{18} examined the Penrose process near
Kerr-Newman-Tab-NUT BH and presented a relation between gained
energy and the corresponding NUT parameter. Mukherjee \cite{19}
explored the collisional Penrose process using spinning particles
around Kerr BH and found that the energy extraction is high in
comparison with the non-spinning case.

The efficiency of energy extraction from a rotating BH by Penrose
process is defined as (gain in energy)/(input energy). Bhat et al.
\cite{20} showed that the efficiency of Penrose process near
Kerr-Newman BH gets reduced due to the vicinity of charge in
comparison with the the maximum efficiency limit of $20.7\%$ for the
Kerr BH. Liu et al. \cite{21} deduced that the deformation parameter
increase the efficiency of energy extraction process in the non-Kerr
BH. Toshmatov et al. \cite{22} studied Penrose process around
rotating regular BH and found that efficiency of energy extraction
decreases for increasing values of electric charge. Liu and Liu
\cite{23} explored the Penrose process using spinning particles
around extremal Kerr BH and concluded that efficiency of the energy
extraction monotonously increases with the increase of particles
spin.

In this paper, we explore the circular geodesic motion and energy
extraction by Penrose process in the background of Kerr-MOG BH. In
section $\textbf{2}$, we introduce the Kerr-MOG BH and study the
null geodesics. We derive the equation for circular photon orbits
and discuss the stability of these orbits through effective
potential. The Lyapunov exponent and the effective force acting on
photons is also discussed. Section $\textbf{3}$ is devoted to
investigate the negative energy state as well as the efficiency of
energy extraction via Penrose process. Finally, we summarize our
results in last section.

\section{Equatorial Circular Geodesics}

In this section, we explore the geodesic motion of neutral particles
near Kerr-MOG BH which can be characterized by the angular momentum $J=Ma$, mass $M$ as well as
the dimensionless parameter $\alpha$ \cite{24}. The spacetime
geometry corresponding to this BH can be described by the metric
with Boyer-Linquist coordinates
\begin{eqnarray}\nonumber
ds^{2}&=&-\frac{\Delta}{\rho^{2}}(dt-a\sin^{2}\theta
d\phi)^{2}+\frac{\sin^{2}\theta}{\rho^{2}}\left[(r^{2} +a^{2})d\phi
- adt\right]^{2}\\\label{1}&
+&\frac{\rho^{2}}{\Delta}dr^{2}+\rho^{2} d\theta^{2},
\end{eqnarray}
with
\begin{eqnarray}\nonumber
g_{tt}&=&-\left(\frac{\Delta-a^{2}\sin^{2}\theta}{\rho^{2}}\right),
\quad g_{rr}=\frac{\rho^{2}}{\Delta}, \quad
g_{\theta\theta}=\rho^{2}, \\\nonumber
g_{\phi\phi}&=&\frac{\sin^{2}\theta}{\rho^{2}}\left[(r^{2}+a^{2})^{2}-\Delta
a^{2}\sin^{2}\theta\right], \\\nonumber
g_{t\phi}&=&\frac{a\sin^{2}\theta}{\rho^{2}}\left[\Delta-(r^{2}+a^{2})\right].
\end{eqnarray}
where
\begin{equation}\nonumber
\Delta=r^{2}+a^{2}-2GMr+\alpha G_{N}GM^{2},\quad
\rho^{2}=r^{2}+a^{2}\cos^{2}\theta,
\end{equation}
here $G=G_{N}(1+\alpha)$ is the gravitational constant, $G_{N}$ is
the Newton's gravitational constant. Moreover, $\alpha$ is
introduced to analyze the strength of the gravitational interaction.
It is worthy to state that the gravitational source charge of the (spin 1 graviton)
vector field $\phi_\mu$ is $Q_g=\sqrt{\alpha G_N}M$.
Inserting the value of $G$ in $\Delta$ and using $r\rightarrow
r/G_{N}M$, $a\rightarrow a/G_{N}M$ and $\alpha\rightarrow
\alpha/G_{N}M$, we obtain
\begin{equation}\nonumber
\Delta=r^{2}+a^{2}-2(1+\alpha)r+\alpha(1+\alpha).
\end{equation}
The Kerr-MOG BH is asymptotically flat, axially and stationary
symmetric solution of MOG field equations. This has the form of a
Kerr-Newman solution of Einstein-Maxwell equations that illustrates
the geometry of a spacetime in the background of a rotating and
electrically charged body. The metric of Kerr-MOG BH is identical to
Kerr-Newman BH with replacement of $\alpha$ by $Q^{2}$, where $Q$ is
the electric charge of Kerr-Newman BH. For $\alpha=0$ and $a=0$, the
metric (\ref{1}) boils down to the Kerr and Schwarzschild-MOG
metric, respectively. Moreover, the zero choices of both of these
parameters describes the Schwarzschild spacetime. On substituting
$\Delta=0$, the horizons of Eq.(\ref{1}) are found to be
\begin{equation}\nonumber
r_{\pm}=(1+\alpha)\pm\sqrt{1+\alpha-a^{2}},
\end{equation}
where $\pm$ sign corresponds to the inner and outer horizons,
respectively. Figure $\textbf{1}$ depicts the horizon structure of
Kerr-MOG, Kerr and Schwarzschild-MOG BH. The left graph indicates
that the horizon of Kerr-MOG BH is greater than that of Kerr BH.
Furthermore, Kerr-MOG BH horizon increases in size with the
increasing value of parameter $\alpha$. The right graph shows that
the horizon of Schwarzschild-MOG BH is larger in comparison with the
Kerr-MOG BH.
\begin{figure*}
\begin{minipage}[b]{0.53\textwidth}
\includegraphics[width=0.9\textwidth]{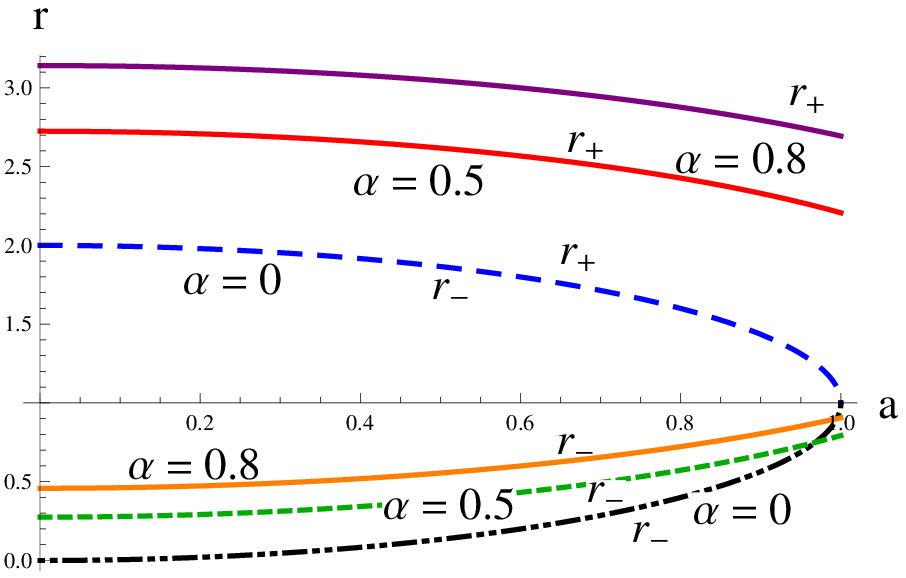}
\end{minipage}\hspace{-0.4cm}
\begin{minipage}[b]{0.53\textwidth}
\includegraphics[width=0.9\textwidth]{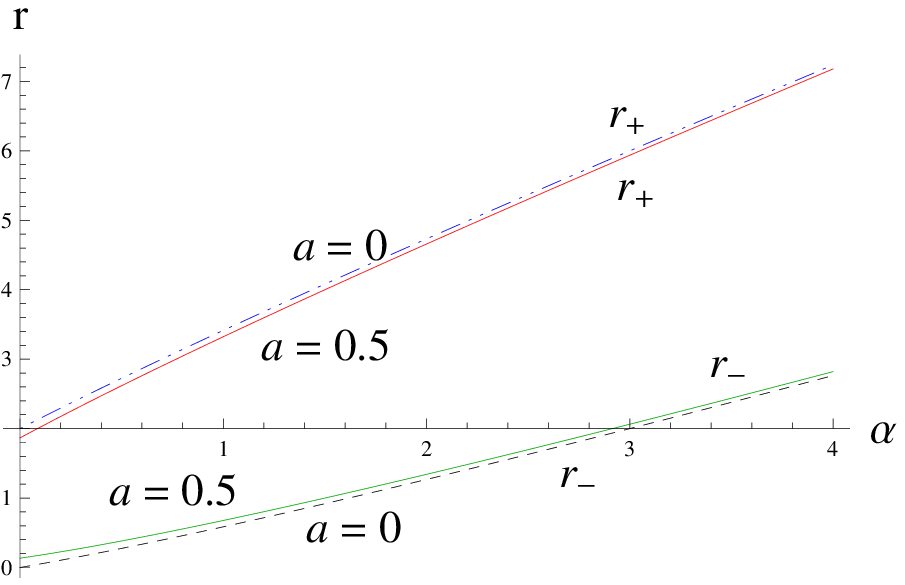}
\end{minipage}
\caption{Plots of horizon as a function of $a$ (left) and $\alpha$
(right).}\label{fig:1}
\end{figure*}
The ergosphere can be calculated by $g_{tt}=0$ as
\begin{equation}\nonumber
r_{es}=\frac{1}{2}[2(1+\alpha)\pm\sqrt{4-2a^{2}+4\alpha-2a^{2}\cos2\theta}].
\end{equation}
\begin{figure*}
\vspace{0.7cm}
    \begin{minipage}[b]{0.53\textwidth}
        \includegraphics[width=0.9\textwidth]{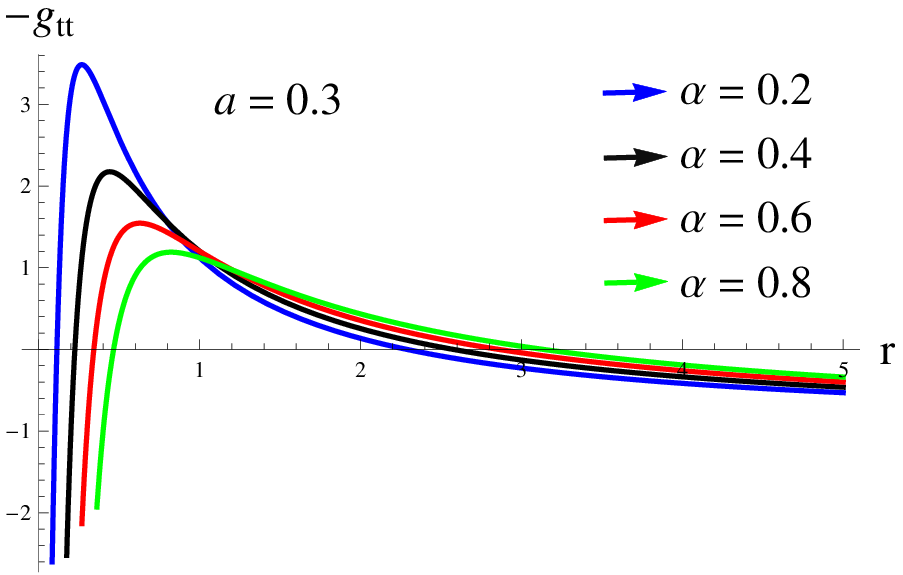}
    \end{minipage}\hspace{-0.4cm}
    \begin{minipage}[b]{0.53\textwidth}
        \includegraphics[width=0.9\textwidth]{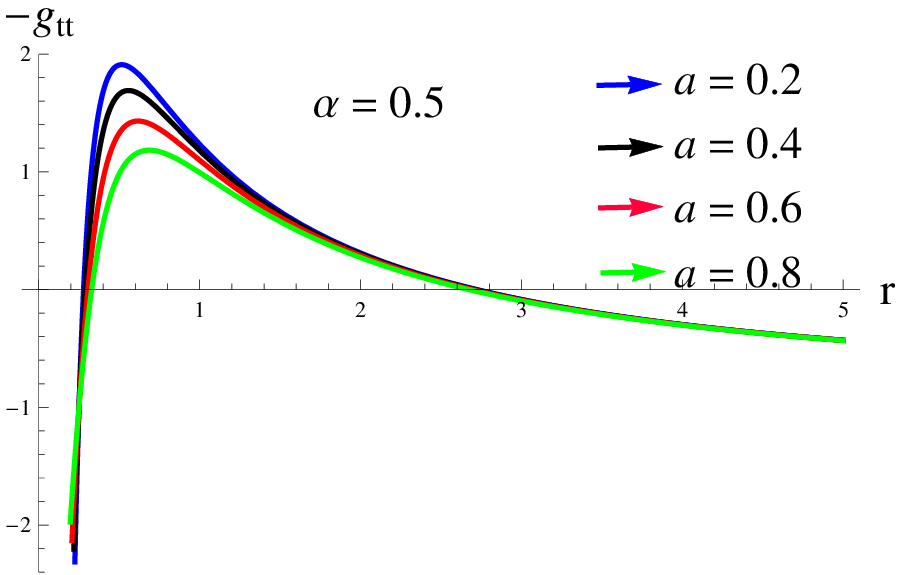}
    \end{minipage}
    \caption{Plots of $-g_{tt}$ as a function of $r$.}\label{fig:2}
\end{figure*}
For $\theta=0, \pi$, both event horizon and ergosphere coincide.
Figure $\textbf{2}$ illustrates the nature of the variation of
$-g_{tt}$ with respect to the radial coordinate $r$. We observe that
the shape of the ergoregion gets modified for increasing values of
the spin parameter $a$ as well as $\alpha$ and becomes constant at a
larger radial distance $r$.

We consider the circular motion of test particles in the background of
Kerr-MOG BH and restrict ourselves to the case of orbits situated on
an equatorial plane ($\theta=\pi/2$). The test particle action can
be written as \cite{1}
\begin{eqnarray}\label{m1}
S=-\left(m\int d\tau+\lambda\int\omega
d\tau\phi_{\mu}\frac{dx^{\mu}}{d\tau}\right),
\end{eqnarray}
where $m$ is the mass of the particle, $\lambda$ is the coupling
constant and $\tau$ is the proper time of the test particle. Using
stationary condition $\delta S/\delta x^{\mu}=0$ in the above
equation, the equation of motion for massive particle can be found
as \cite{1a}
\begin{eqnarray}\label{m2}
\frac{d^{2}x^{\mu}}{d\tau^{2}}+\Gamma^{\mu}_{\nu\beta}\frac{dx^{\nu}}
{dx^{\tau}}\frac{dx^{\beta}}{dx^{\tau}}=
\frac{q}{m}B^{\mu}_{\sigma}\frac{dx^{\sigma}}{d\tau},
\end{eqnarray}
where $\Gamma^{\mu}_{\nu\beta}$ denotes the Christoffel symbols,
$q=\sqrt{\alpha G_{N}}M$  is the test particle gravitational charge and
$B_{\mu\sigma}=\partial_{\mu}\phi_{\sigma}-\partial_{\sigma}\phi_{\mu}$.
The equation of motion for the photon is
\begin{eqnarray}\label{m3}
\frac{d^{2}x^{\mu}}{d\tau^{2}}+\Gamma^{\mu}_{\nu\beta}\frac{dx^{\nu}}
{dx^{\tau}}\frac{dx^{\beta}}{dx^{\tau}}= 0.
\end{eqnarray}
The radial equation of motion for a test particle in an equatorial
plane for Kerr-MOG BH can be written as \cite{1b}
\begin{eqnarray}\label{8}
A(r,\alpha)E^{2}-2B(r,L,\alpha)E+C(r,L,\alpha)+r^{4}(p^{r})^{2}=0,
\end{eqnarray}
where $p^{r}$ is the radial momentum, $E$ and $L$ denotes the energy
and angular momentum of test particle, and
\begin{eqnarray}\nonumber
A(r,\alpha)&=&(r^{2}+a^{2})^{2}-\Delta^{2}a^{2},\\\nonumber
B(r,L,\alpha)&=&(La+r\alpha)(r^{2}+a^{2})-La\Delta,\\\nonumber
C(r,L,\alpha)&=&(La+r\alpha)^{2}-L^{2}\Delta-r^{2}\Delta.
\end{eqnarray}
In the following, we discuss the null geodesics, stability of
circular photons orbits, effective force and Lyapunov exponent.

\subsection{Null Geodesics}

For null geodesics, the radial equation can be written as
\begin{eqnarray}\nonumber
\dot{r}^{2}&=&E^{2}+\frac{1}{r^{4}}(2(1+\alpha)r-\alpha
(1+\alpha))(aE-L)^{2}\\\label{9}&+&\frac{1}{r^{2}}(a^{2}E^{2}-L^{2}),
\end{eqnarray}
where dot denotes the derivative with respect to $\tau$. It could be
worthwhile to introduce the impact parameter $D=L/E$ instead of $L$.
Now, we consider a particular choice, i.e., $L=aE$ for which $D=a$.
Thus, the equations for $\dot{t}$, $\dot{\phi}$ and $\dot{r}$ can be
written as
\begin{eqnarray}\nonumber
\dot{t}=\frac{E\left(r^{2}+a^{2}\right)}{\Delta}, \quad
\dot{\phi}=\frac{Ea}{\Delta}, \quad \dot{r}=\pm E,
\end{eqnarray}
where $\pm$ sign correspond to the outgoing and incoming photons,
respectively. The radial coordinate $r$ is specified with respect to
affine parameter while the equations for $t$ and $\phi$ are
\begin{eqnarray}\nonumber
\frac{dt}{dr}=\pm\frac{\left(r^{2}+a^{2}\right)}{\Delta}, \quad
\frac{d\phi}{dt}=\pm\frac{a}{\Delta}.
\end{eqnarray}
Now, we examine the general case $L\neq aE$ and determine the radius
of unstable photon orbit for which $E=E_{c}$, $L=L_{c}$ and
$D_{c}=L_{c}/E_{c}$. Therefore, Eq.(\ref{9}) and its derivative
takes the form
\begin{eqnarray}\nonumber
r_{c}^{2}&+&\left(\frac{2(1+\alpha)r_{c}-\alpha
(1+\alpha)}{r_{c}^{2}}\right)\left(a-D_{c}\right)^{2}\\\label{10}&
+&\left(a^{2}-D_{c}^{2}\right)=0,
\end{eqnarray}
\begin{eqnarray}\label{11}
r_{c}-\left(\frac{(1+\alpha)r^{2}_{c}-\alpha
(1+\alpha)r_{c}}{r_{c}^{4}}\right)\left(a-D_{c}\right)^{2}=0.
\end{eqnarray}
The last equation implies that
\begin{eqnarray}\label{12}
D_{c}=a\mp\sqrt{\frac{r_{c}^{5}}{(1+\alpha)r^{2}_{c}-\alpha
(1+\alpha)r_{c}}}.
\end{eqnarray}
Substituting Eq.(\ref{12}) into Eq.(\ref{10}), we obtain
\begin{eqnarray}\nonumber
r_{c}^{3}&-&3(1+\alpha)r^{2}_{c}+2\alpha (1+\alpha)r_{c}\\\label{13}&\pm&
2a\sqrt{r_{c}((1+\alpha)r^{2}_{c}-\alpha (1+\alpha)r_{c})}=0.
\end{eqnarray}
\begin{figure*}
    \begin{minipage}[b]{0.53\textwidth}
        \includegraphics[width=0.9\textwidth]{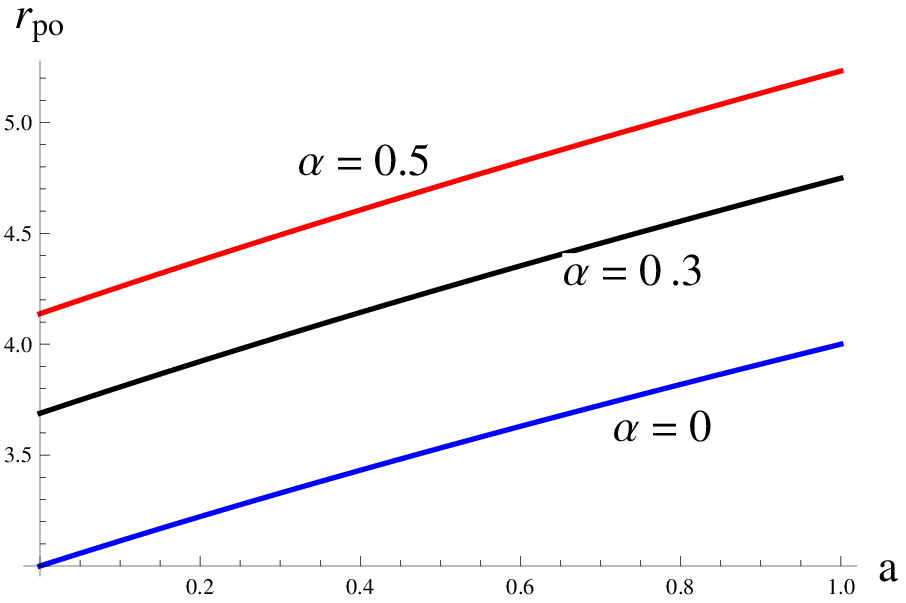}
    \end{minipage}\vspace{-0.4cm}
    \begin{minipage}[b]{0.53\textwidth}
        \includegraphics[width=0.9\textwidth]{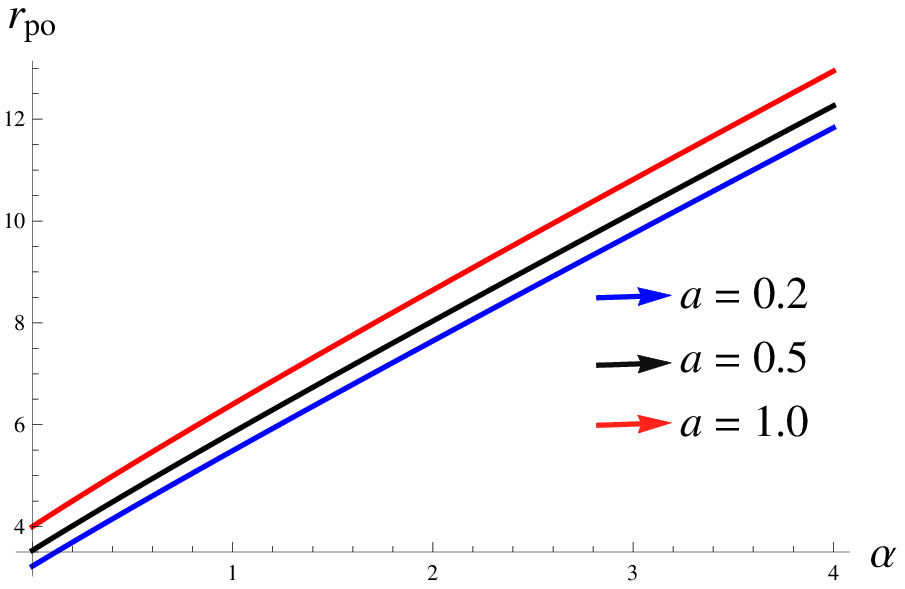}
    \end{minipage}\vspace{0.7cm}
    \caption{Plots of photon
orbit for retrograde motion as a function of $a$ (left) and $\alpha$
(right).}\label{fig:3}
\end{figure*}
\begin{figure*}
    \begin{minipage}[b]{0.53\textwidth}
        \includegraphics[width=0.9\textwidth]{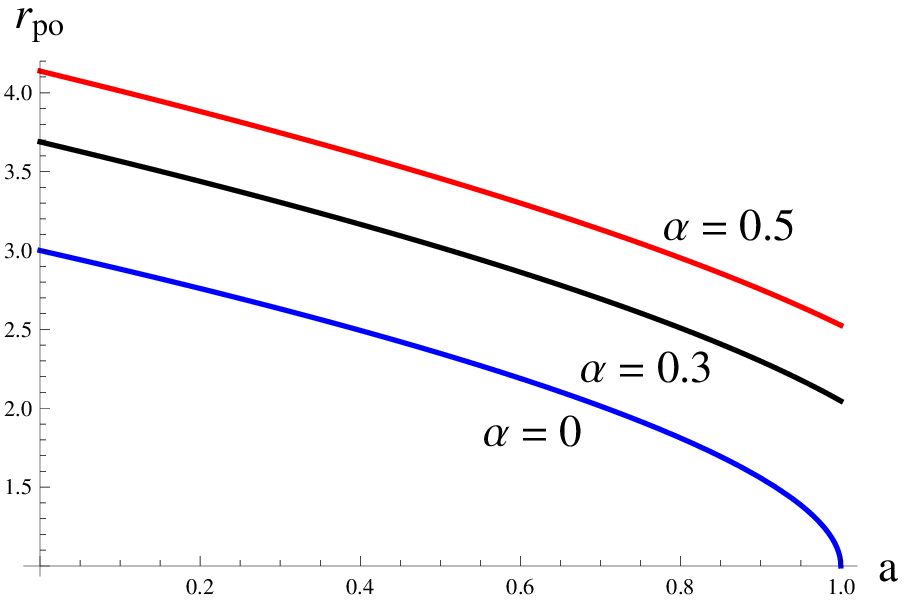}
    \end{minipage}\vspace{-0.4cm}
    \begin{minipage}[b]{0.53\textwidth}
        \includegraphics[width=0.9\textwidth]{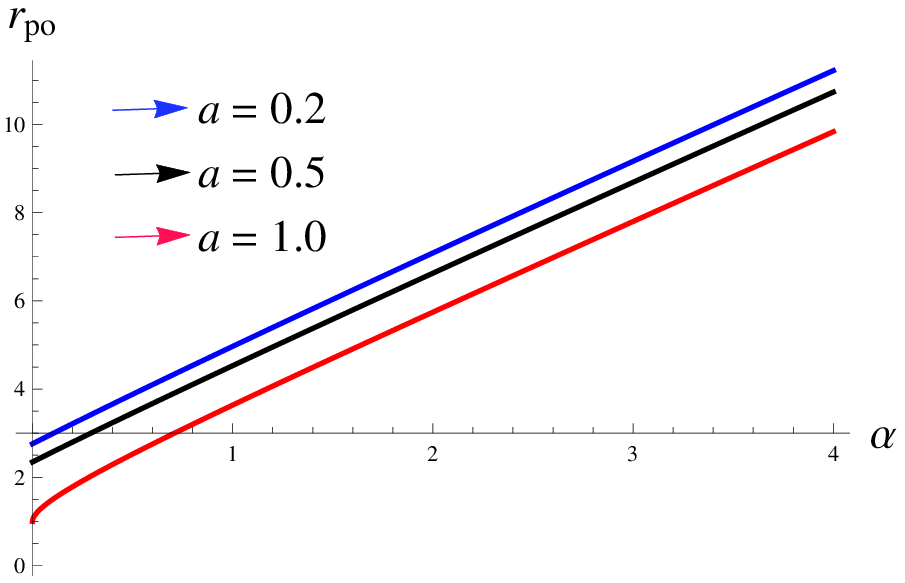}
    \end{minipage}\vspace{0.7cm}
    \caption{Plots of photon
orbit for prograde motion as a function of $a$ (left) and $\alpha$
(right).}\label{fig:3}
\end{figure*}
Let $r_{c}=r_{po}$ be the real positive root of Eq.(\ref{13}) which
gives radius of the photon orbit for Kerr-MOG BH. It is the closest
possible boundary of circular orbits for particles. For
$\alpha=a=0$, we obtain Schwarzschild limit ($r_{po}=3M$), while for
$\alpha=0$ and $a=1$, we recover the photon orbit of Kerr BH
($r_{po}=M$) \cite{9}. The behavior of photon orbit for different
values of $a$ (left) and $\alpha$ (right) for retrograde and
prograde motion is depicted in Figures $\textbf{3}$ and
$\textbf{4}$, respectively. Here, we observe that the radius of
photon orbit of Kerr-MOG BH is greater in comparison with the Kerr
BH and increases for increasing values of $\alpha$ as well as $a$
for retrograde motion. But for prograde motion, the radius of photon
orbit increases with the increase of $\alpha$ and decreases as the
rotation of a BH increases. From Eqs.(\ref{10}) and (\ref{12}), we
have
\begin{eqnarray}\nonumber
D_{c}^{2}=a^{2}+r_{c}^{2}\left[\frac{3(1+\alpha)r^{2}_{c}-2\alpha
(1+\alpha)r_{c}}{(1+\alpha)r^{2}_{c}-\alpha(1+\alpha)r_{c}}\right].
\end{eqnarray}
We can obtain the angular velocity (physical quantity related with
the null circular geodesics) measured by asymptotic observer denoted
by $\Omega_{c}$ as
\begin{eqnarray}\nonumber
\Omega_{c}&=&[(r_{c}^{2}-2(1+\alpha)r_{c}+\alpha
(1+\alpha))D_{c}+(2(1+\alpha)r_{c}\\\nonumber&-&\alpha
(1+\alpha))a][(r_{c}^{2}(r_{c}^{2}+a^{2})+a^{2}(2(1+\alpha)r_{c}-\alpha\\\nonumber&\times&
(1+\alpha)))-a(2(1+\alpha)r_{c}-\alpha
(1+\alpha))D_{c}]^{-1}\\\nonumber&=&\frac{1}{D_{c}}.
\end{eqnarray}
We can see from Eqs.(\ref{10}) and (\ref{12}) that the angular
frequency of null geodesics is inverse of the impact parameter,
which generalize the results of Kerr BH \cite{8} to the Kerr-MOG BH.
\begin{figure*}\vspace{0.7cm}
    \begin{minipage}[b]{0.53\textwidth}
        \includegraphics[width=0.9\textwidth]{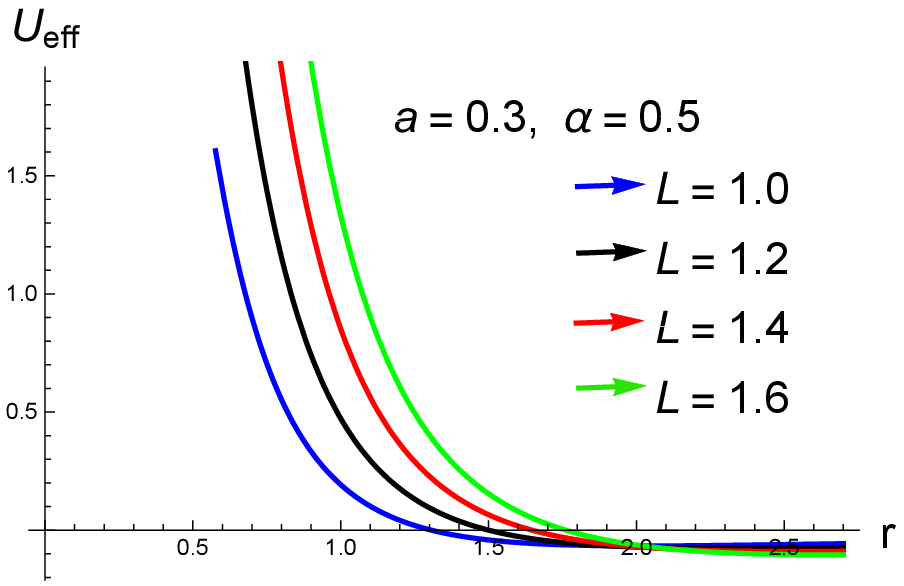}
    \end{minipage} \hspace{-0.4cm}
        \begin{minipage}[b]{0.53\textwidth}
        \includegraphics[width=0.9\textwidth]{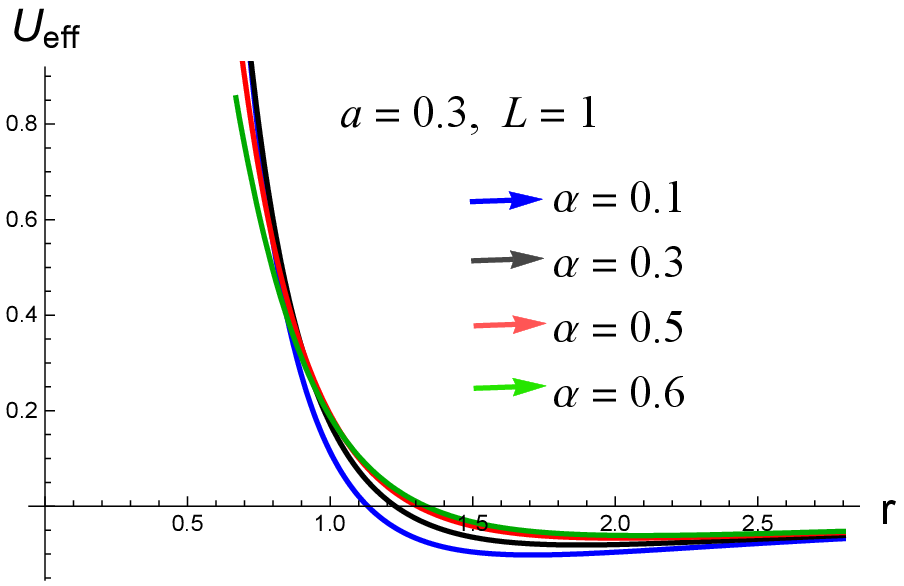}
    \end{minipage}\vspace{0.5cm}
    \begin{minipage}[b]{0.53\textwidth}
        \includegraphics[width=0.9\textwidth]{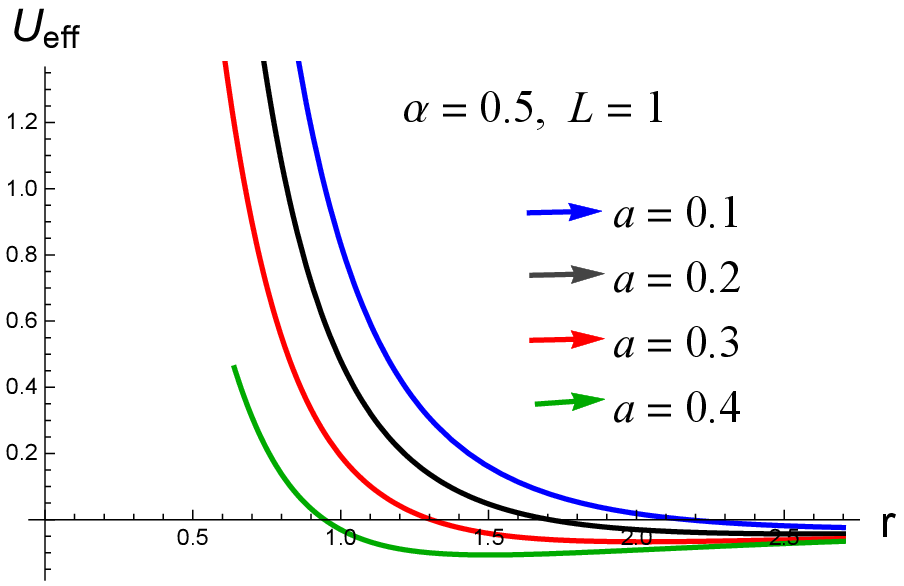}
    \end{minipage}\hspace{-0.4cm}
    \begin{minipage}[b]{0.53\textwidth}
        \includegraphics[width=0.9\textwidth]{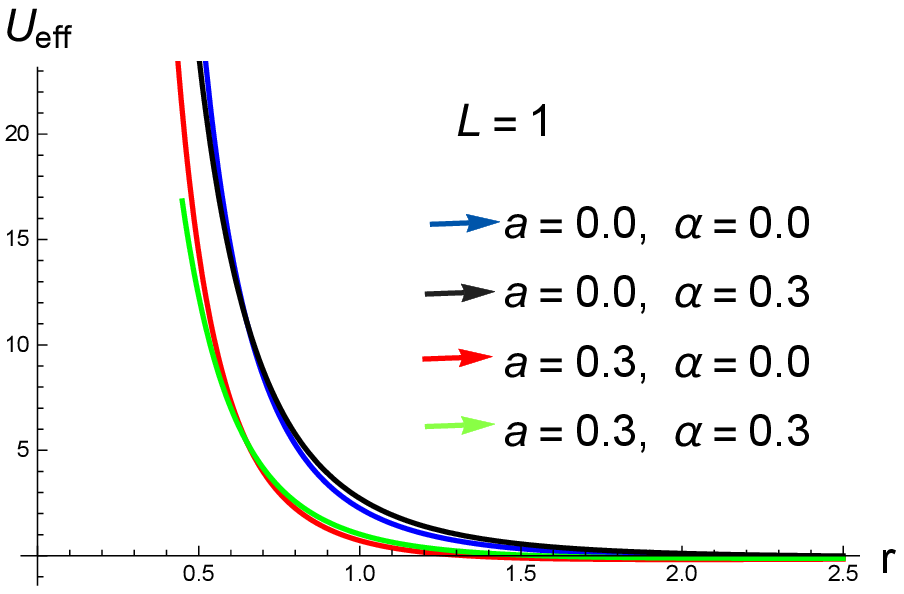}
    \end{minipage}
    \caption{Plots of effective potential versus $r$.}\label{fig:5}
\end{figure*}

The stability of circular orbits can be illustrated through
effective potential ($U_{eff}$). The minimum and maximum values of
$U_{eff}$ determine the stable and unstable circular orbits,
respectively, while the extreme values of $U_{eff}$ correspond to
$U_{eff},r=0$. Equation (\ref{9}) can be written as \cite{24a}
\begin{eqnarray}\nonumber
\dot{r}^{2}&=&E^{2}-U_{eff},
\end{eqnarray}
where
\begin{eqnarray}\nonumber
U_{eff}&=&\frac{1}{r^{4}}\left(\alpha
(1+\alpha)-2(1+\alpha)r\right)\left(a
E-L\right)^{2}\\\label{14}&-&\frac{1}{r^{2}}(a^{2}E^{2}-L^{2}).
\end{eqnarray}
For a particle to describe a circular orbit (at a constant radius
$r=r_{0}$), the initial radial velocity and radial acceleration must
be zero. Therefore, we have
\begin{eqnarray}\nonumber
U_{eff}|_{r=r_{0}}, \quad \frac{\partial
U_{eff}}{\partial r}|_{r=r_{0}}=0.
\end{eqnarray}
For stable circular orbits, effective potential must be minimum,
i.e.,
\begin{eqnarray}\nonumber
\frac{\partial^{2} U_{eff}}{\partial r^{2}}|_{r=r_{0}}>0.
\end{eqnarray}
The behavior of effective potential for null-like particles moving
around Kerr-MOG BH is shown in Figure $\textbf{5}$. We have varied
different choices of angular momentum in this structure and shown
our results in the upper portion of the left graphs. It is noticed
that the instability of circular photon orbits increases with the
increase of angular momentum and at a larger radial distance, does
not change much more. The right graph presents that the presence of
parameter $\alpha$ contributes to decrease the stability of circular
photons orbits. The underneath part of the left diagram indicates
more unstable photons circular orbits for large choices of $a$ as
compared to smaller ones. The motion of particles become unstable
for higher values of $a$. The photons orbits around Kerr-MOG BH are
unstable in comparison with the Kerr BH as shown in the right
diagram.

\subsection{Effective Force}

The force that could describe the motion of information, e.g.,
whether it is directed away or attracted towards the BH, could
termed as an effective force. One can obtain effective force acting
on the photon using Eq.(\ref{14}) as \cite{13}
\begin{eqnarray}\nonumber
F&=&\frac{-1}{2}\left(\frac{dU_{eff}}{dr}\right)\\\nonumber
&=&-\frac{(1+\alpha)}{r^{4}}(a
E-L)^{2}+\frac{1}{r^{3}}(a^{2}E^{2}-L^{2})\\\nonumber
&+&\frac{2(a E-L)^{2}}{r^{5}}\left(2r(1+\alpha)-\alpha
(1+\alpha)\right).
\end{eqnarray}
We see that the first term is attractive while the second term is repulsive if $aE>L$. The third term
is also repulsive if $2r(1+\alpha)>\alpha(1+\alpha)$. The
graphical analysis of effective force is analyzed in Figure
$\textbf{6}$. The top portion of the left diagram presents
relatively greater attraction on photons by the effective force with
higher choices of spin parameter $a$ and at a larger radial distance
does not change much more as $a$ increases. The behavior of the
effective force with various $\alpha$ choices is depicted in the
right diagram. This states the increment on the influence of
effective force with respect to the increasing $\alpha$ values and
has no effects as the photons move away from the BH. The lower graph
gives the comparison of the effective force acting on a photon
around the Kerr-MOG BH with the Schwarzschild and Kerr structures.
One can observe the more attractive behavior of effective force on a
photon for Kerr-MOG BH than that of Schwarzschild and Kerr BHs.
\begin{figure*}
    \begin{minipage}[b]{0.53\textwidth}
        \includegraphics[width=0.9\textwidth]{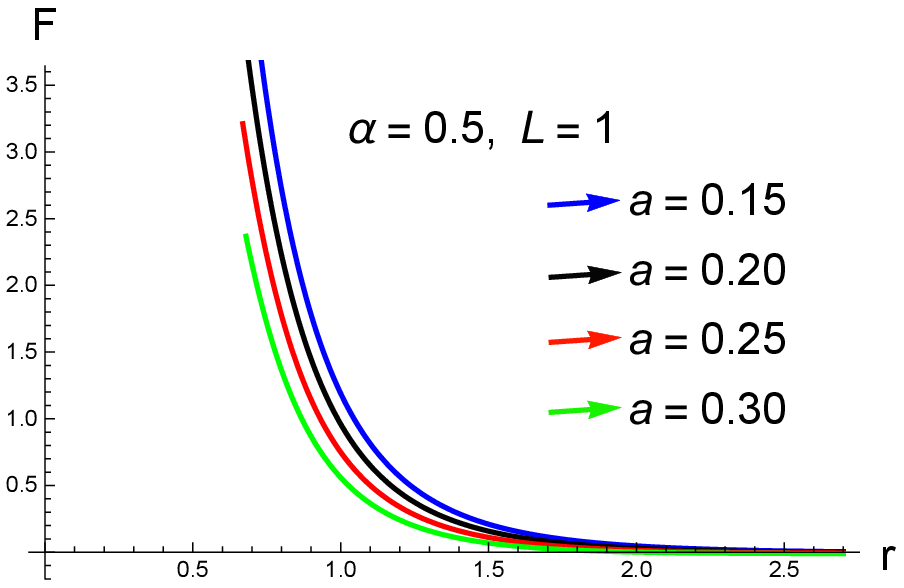}
    \end{minipage}\hspace{-0.4cm}
    \begin{minipage}[b]{0.53\textwidth}
        \includegraphics[width=0.9\textwidth]{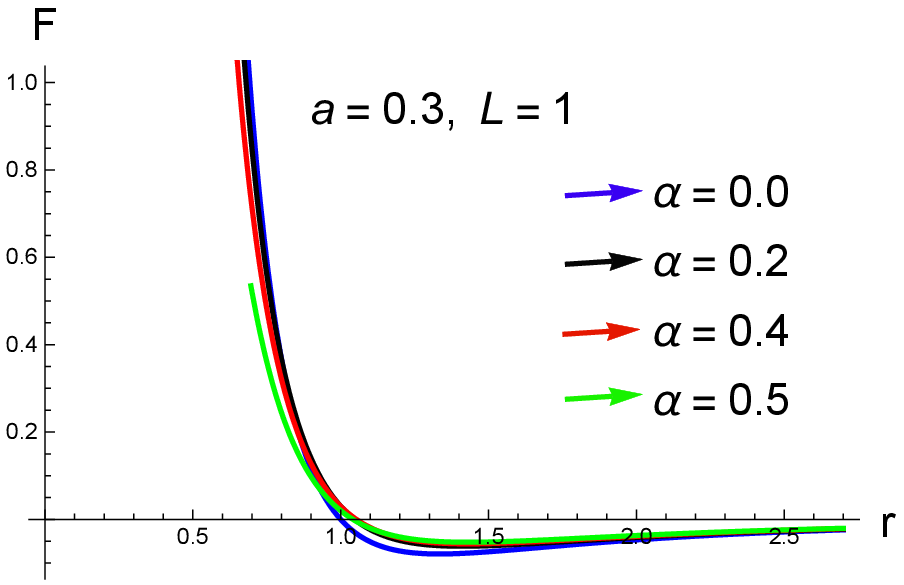}
    \end{minipage}
    \vspace{0.7cm}
    \center
    \begin{minipage}[b]{0.53\textwidth}
        \includegraphics[width=0.9\textwidth]{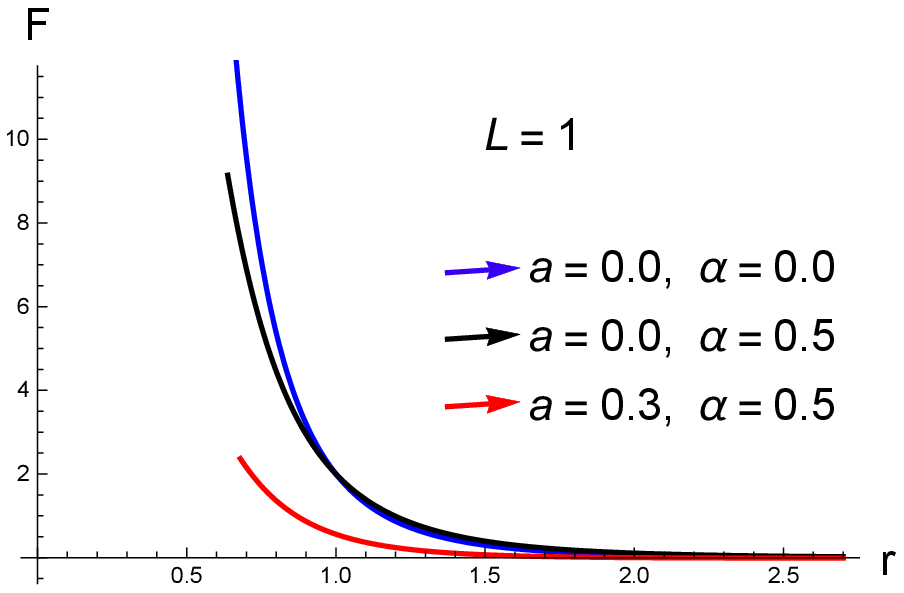}
    \end{minipage}
    \caption{Plots of effective force versus $r$.}\label{fig:6}
\end{figure*}

\subsection{Lyapunov Exponent}

\begin{figure*}
    \begin{minipage}[b]{0.53\textwidth}
        \includegraphics[width=0.9\textwidth]{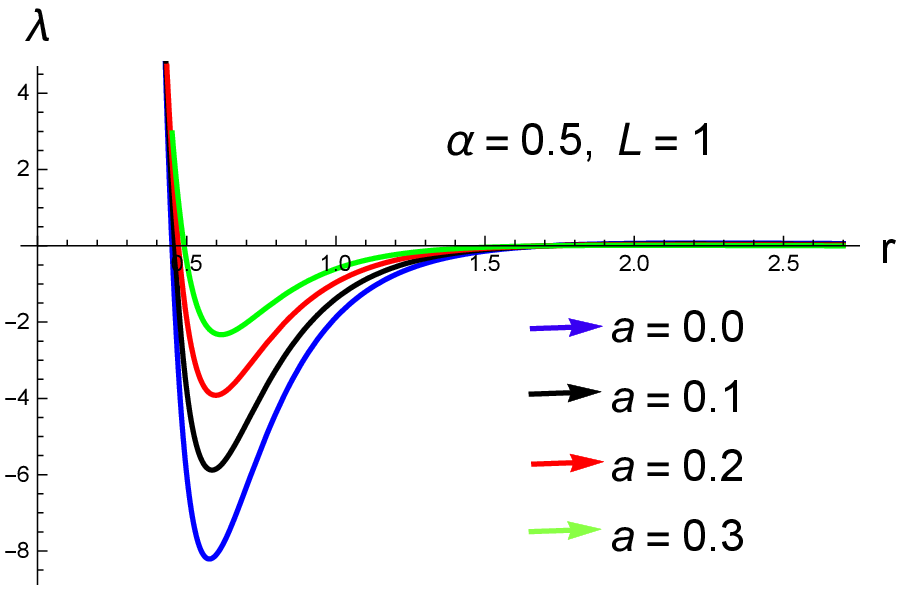}
    \end{minipage}\hspace{-0.4cm}
    \begin{minipage}[b]{0.53\textwidth}
        \includegraphics[width=0.9\textwidth]{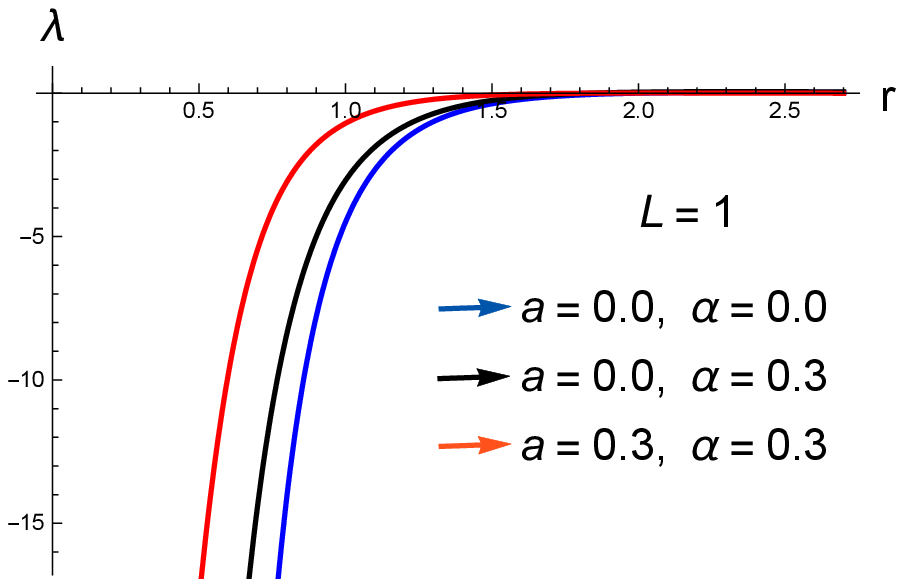}
    \end{minipage}
    \center
    \vspace{0.7cm}
    \begin{minipage}[b]{0.53\textwidth}
        \includegraphics[width=0.9\textwidth]{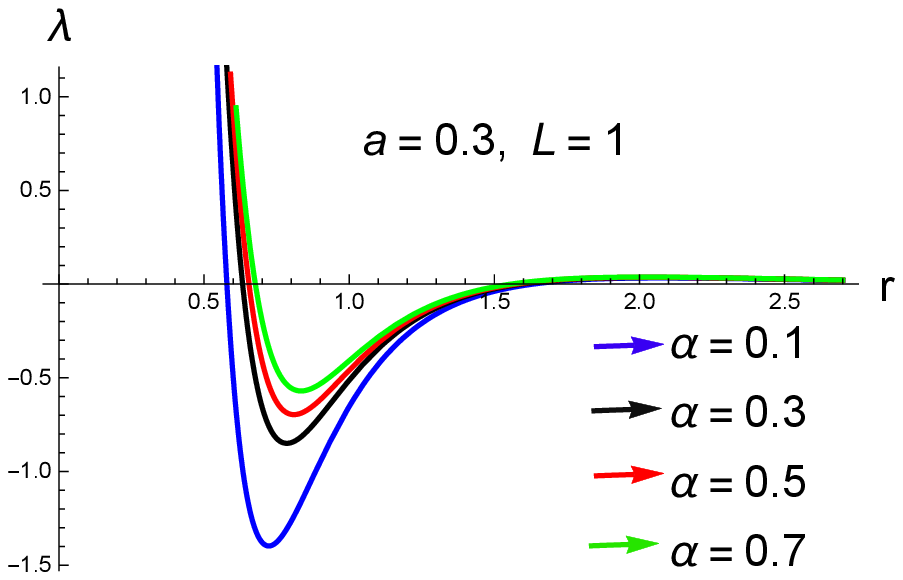}
    \end{minipage}
    \caption{Plots of Lyapunov exponent versus $r$.}\label{fig:7}
\end{figure*}
The average contraction and expansion geodesics rates can be calculated in a phase space
through Lyapunov exponent. The negative and positive
Lyapunov exponents determines the convergence and divergence between
nearby orbits. After making use of Eq.(\ref{14}), the Lyapunov exponent
for null geodesics has been found to be \cite{25}
\begin{eqnarray}\nonumber
\lambda&=&\sqrt{\frac{-U_{eff}^{\prime\prime}(r_{c})}{2\dot{t}^{2}(r_{c})}}\\\nonumber
&=&[\frac{1}{2L^{2}r^{6}}(a E-L)(\alpha+(r-2)r)
(10\alpha^2(L-a E)\\\nonumber&+&2\alpha(6r-5)(a E-L)
+3r(a E(r+4)\\\nonumber&+&L(r-4)))]^{\frac{1}{2}}\mid_{r=r_{c}}.
\end{eqnarray}
In the upper portion of Figure $\textbf{7}$, the left diagram
describes more unstable circular orbits for large choices of $a$,
which indicates the decreasing behavior of Lyapunov exponent with
bigger selection of $a$. However, the right diagram indicates that
the Lyapunov exponent for Kerr-MOG BH has larger values as compared
to Schwarzschild and Schwarzschild-MOG structures. We have plotted
lower portion of the diagram for various distinct choices of
$\alpha$. We noticed that initially $\lambda$ has decreasing
behavior for higher values of $\alpha$ and then increases with the
increase of $r$ but does not change much more at a larger radial
distance.

\section{Energy Extraction by Penrose Process}

The energy extraction process from BHs is the momentous problem of
general relativity. There are different mechanisms which are devoted
to study the energy extraction from rotating BHs. Among these
processes, the Penrose process is of great interest and even more
efficient than the nuclear reactions. In this process, a particle
with positive energy enters into the ergosphere (the region between
outer horizon and stationary limit), splitted into two fragments,
one of them follows a negative energy orbit while the other escapes
to infinity having energy greater than that of the incident
particle. If the particles are involved in Penrose process then the
necessary and sufficient condition to extract energy is the
absorption of particles with negative energies as well as angular
momentum. In the following, we study the negative energy state as
well as efficiency of energy extraction by the Penrose process
around Kerr-MOG BH.

\subsection{The Negative Energy State}

\begin{figure*}\vspace{0.7cm}
        \begin{minipage}[b]{0.53\textwidth}
        \includegraphics[width=0.9\textwidth]{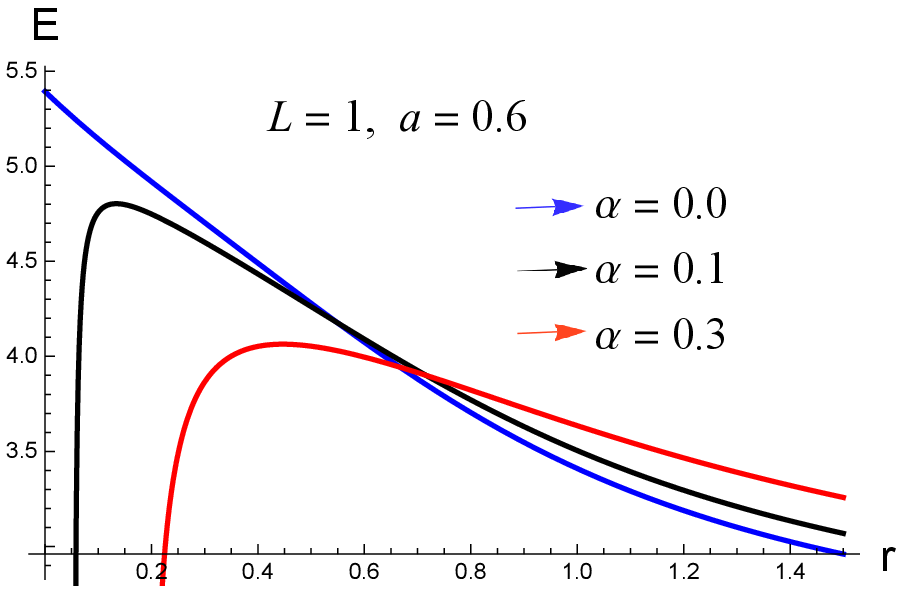}
    \end{minipage}\hspace{-0.4cm}
    \begin{minipage}[b]{0.53\textwidth}
        \includegraphics[width=0.9\textwidth]{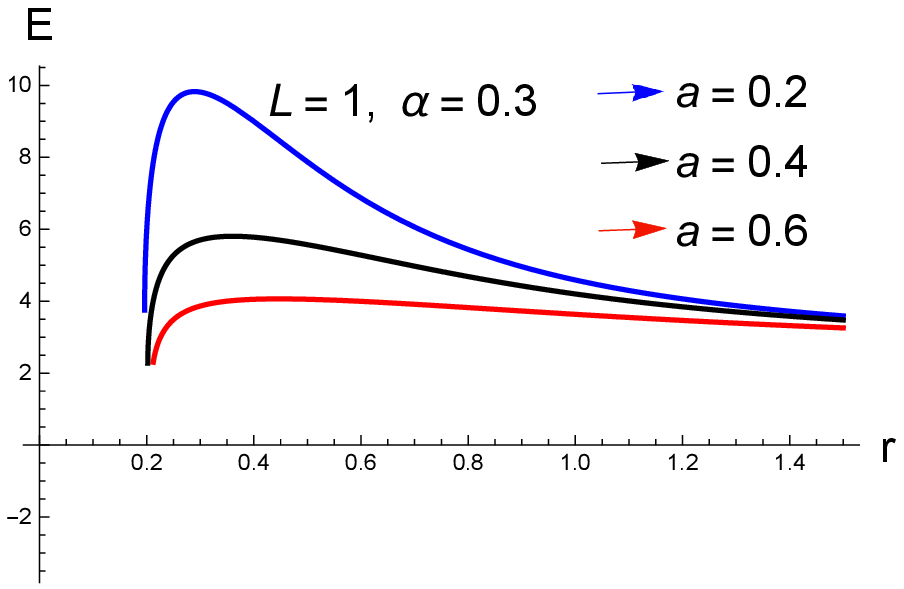}
    \end{minipage}
    \caption{Plots of negative energy as a function of $\alpha$ (left) and $a$ (right).}\label{fig:8}
\end{figure*}

The negative energy inside the ergosphere has an important
consequences in BH physics. It may allows the Penrose process and
occur due to the electromagnetic interaction and the counter
rotating orbits around BH. Thus, it is useful to find the limits on
energy which a particle can have at a particular location. Using
Eq.(\ref{8}), we have
\begin{eqnarray}\nonumber
E^{2}[r^{2}(r^{2}&+&a^{2})+a^{2}(2(1+\alpha)r-\alpha
(1+\alpha))]-2E( a
L(2(1+\alpha)r\\\nonumber&-&\alpha(1+\alpha))+\alpha
r(r^{2}+a^{2}))-L^{2}(r^{2}-2(1+\alpha)r\\\label{15}&+&\alpha
(1+\alpha))+\alpha r(\alpha r+2aL)-r^{2}\Delta=0.
\end{eqnarray}
Solving the above equation for $E$ and $L$, we obtain
\begin{eqnarray}\label{16}
E&=&\frac{a L(2(1+\alpha)r-\alpha(1+\alpha)) \pm \sqrt{X_{1}} }
{r^{4}+a^{2}(r^{2}+2(1+\alpha)r-\alpha(1+\alpha))},
\\\label{17} L&=&\frac{-a E(2(1+\alpha)r-\alpha(1+\alpha))\pm \sqrt{X_{2}}
} {(r^{2}-2(1+\alpha)r+\alpha(1+\alpha))},
\end{eqnarray}
where
\begin{eqnarray}\nonumber
X_{1}&=&-a^2\Delta^2(4a^2+3L^2)-3(a^2+r^2)^2(a L+\alpha r)^2+2
\Delta(2a^6+a^4(3L^2\\\nonumber&+&4r^2)+3\alpha
a^3Lr+a^2r^2(3L^2+2(\alpha ^2+r^2))-\alpha a Lr^3+2L^2r^4),
\\\nonumber
X_{2}&=&\alpha r(a^2+r^2)+E^2(a^2(-\alpha(\alpha+1)+r^2+2(\alpha+1)
r)+r^4)+4a^2\\\nonumber&\times&(E(\alpha+1)(\alpha-2r)+\alpha
r)^2+\alpha^2+4 \Delta r^2(\alpha^2+\alpha+r^2-2(\alpha+1)r).
\end{eqnarray}
From Eq.(\ref{16}), we can inferred the conditions for which energy
can be negative. Firstly, we assign an energy $E=1$ to the particle
with unit mass, at rest at infinity. We consider $+$ sign of Eq.(\ref{16})
which requires for $E<0$, $L<0$ and
\begin{eqnarray}\nonumber
\left(aL\left(2\left(1+\alpha\right)r-\alpha
\left(1+\alpha\right)\right)+\alpha r(r^{2}+a^2)\right)^{2}>X_{1}.
\end{eqnarray}
Using Eq.(\ref{18}), the above inequality can be written as
\begin{eqnarray}\nonumber
\left[r^{4}+a^{2}\left(r^{2}+2\left(1+\alpha\right)r-\alpha
(1+\alpha)\right)\right][(r^{2}\\\nonumber-2(1+\alpha)r+\alpha(1+\alpha))L^{2}-\epsilon
r^{2}\Delta]<0.
\end{eqnarray}
It follows from the above inequality that $E<0$ if and only if $L<0$
and
\begin{eqnarray}\nonumber
\left(\frac{r^{2}-2(1+\alpha)r+\alpha
(1+\alpha)}{r^{2}}\right)<\frac{\epsilon \Delta}{L^{2}}.
\end{eqnarray}
We observe that on an equatorial plane, only counter rotating
particles can have negative energy and particle must be inside the
ergosphere ($r<a+M$). In the Penrose process, orbit of particle
having negative energy inside the ergosphere is a key to extract the
energy from Kerr-MOG BH. Figure $\textbf{8}$ describes the negative
energy state for different values of $\alpha$ (left) as well as $a$
(right). It is observed that negative energy decreases with the
increasing values of parameter $\alpha$. It is interesting to note
that negative energy decreases when a BH rotates rapidly.

\subsection{Efficiency of Energy Extraction}

The efficiency of the energy extraction via Penrose process is one
of the most important issues in the energetics of BHs. We assume
that a particle with energy $E_{0}$ enters into the ergosphere of BH
and splitted into two pieces (namely $1$ and $2$ with energies
$E_{1}$ and $E_{2}$, respectively). The piece $1$ has greater energy
than the incident particle and leaves the ergosphere while second
piece with negative energy falls into the BH. According to the law
of conservation of energy
\begin{eqnarray}\nonumber
E_{0}=E_{1}+E_{2},
\end{eqnarray}
where $E_{2}<0$, then $E_{1}>E_{0}$. Let $\Omega=\frac{d\phi}{dt}$
and $\upsilon=\frac{dr}{dt}$ represents the angular and the radial
velocity of a particle with respect to an asymptotic infinity
observer, respectively. From the laws of conservation of energy and
angular momentum, we obtain
\begin{eqnarray}\label{18}
E=-p^{t}\chi, \quad L=p^{t}\Omega,
\end{eqnarray}
where
\begin{eqnarray}\nonumber
\chi\equiv g_{tt}+g_{t\phi}\Omega.
\end{eqnarray}
Using $p^{\sigma}p_{\sigma}=-m^{2}$, we can obtain
\begin{eqnarray}\nonumber
g_{tt}\dot{t}^{2}+2g_{t\phi}\dot{t}\dot{\phi}+g_{rr}\dot{r}^{2}+g_{\phi\phi}\dot{\phi}^{2}=-m^{2}.
\end{eqnarray}
Dividing the above equation by $\dot{t}^{2}$, we have
\begin{eqnarray}\label{18a}
g_{tt}+2g_{t\phi}\Omega+\frac
{r^{2}}{\Delta}\upsilon^{2}+g_{\phi\phi}\Omega^{2}=-\left(\frac{m
\chi}{E}\right)^{2}.
\end{eqnarray}
As the right hand side of above equation is negative or equal to
zero and third term in the left hand side is always positive. So, we
can write the above equation in the following form
\begin{eqnarray}\nonumber
g_{tt}+2g_{t\phi}\Omega+g_{\phi\phi}\Omega^{2}=-\left(\frac{m
\chi}{E}\right)^{2}-\frac {r^{2}}{\Delta}\upsilon^{2}\leq 0.
\end{eqnarray}
The angular velocity $\Omega$ must be in the range of
$\Omega_{-}\leq\Omega\leq\Omega_{+}$ \cite{17}, where
\begin{eqnarray}\nonumber
\Omega_{\pm}=-\frac{g_{t\phi}}{g_{\phi\phi}}
\pm\sqrt{\frac{g^{2}_{t\phi}
-g_{tt}g_{\phi\phi}}{g^{2}_{\phi\phi}}}.
\end{eqnarray}
Using Eq.(\ref{18}), the equations of conservation of energy and
angular momentum can be written as
\begin{eqnarray}\label{19}
p^{t}_{(0)}\chi_{(0)}=p^{t}_{(1)}\chi_{(1)}+p^{t}_{(2)}\chi_{(2)},
\end{eqnarray}
\begin{eqnarray}\label{20}
p^{t}_{(0)}\Omega_{(0)}=p^{t}_{(1)}\Omega_{(1)}+p^{t}_{(2)}\Omega_{(2)}.
\end{eqnarray}
The efficiency $(\eta)$ of the Penrose process is defined as
\begin{eqnarray}\nonumber
\eta=\frac{E_{(1)}-E_{(0)}}{E_{(0)}}=Y-1,
\end{eqnarray}
where $Y=E_{(1)}/E_{(0)}$ and $Y>1$. Using Eqs.(\ref{18}),
(\ref{19}) and (\ref{20}), we obtain
\begin{eqnarray}\label{21}
Y=\frac{E_{(1)}}{E_{(0)}}=\frac{(\Omega_{(0)}-\Omega_{(2)})\chi_{(1)}}
{(\Omega_{(1)}-\Omega_{(2)})\chi_{(0)}}.
\end{eqnarray}
Now, we assume that the incident particle with energy $E_{(0)}=1$
enters into the ergosphere and splitted into two photons having
momenta $p_{(1)}=p_{(2)}=0$. It follows from Eq.(\ref{21}), the
efficiency is maximized if we have the largest value of
$\Omega_{(2)}$ and the smallest value of $\Omega_{(1)}$
simultaneously, which can be obtained when
$\upsilon_{(1)}=\upsilon_{(2)}=0$. In this case
\begin{eqnarray}\label{22}
\Omega_{(1)}=\Omega_{+}, \quad \Omega_{(2)}=\Omega_{-}.
\end{eqnarray}
The corresponding values of parameter $\chi$ are
\begin{eqnarray}\label{23}
\chi_{(0)}=g_{tt}+g_{t\phi}\Omega_{(0)}, \quad
\chi_{(2)}=g_{tt}+g_{t\phi}\Omega_{-}.
\end{eqnarray}
The four-momenta of pieces are
\begin{eqnarray}\nonumber
p_{\sigma}=p^{t}(1,0,0,\Omega_{\sigma}),\quad \sigma=1,2.
\end{eqnarray}
\begin{figure*}
    \begin{minipage}[b]{0.53\textwidth}
        \includegraphics[width=0.9\textwidth]{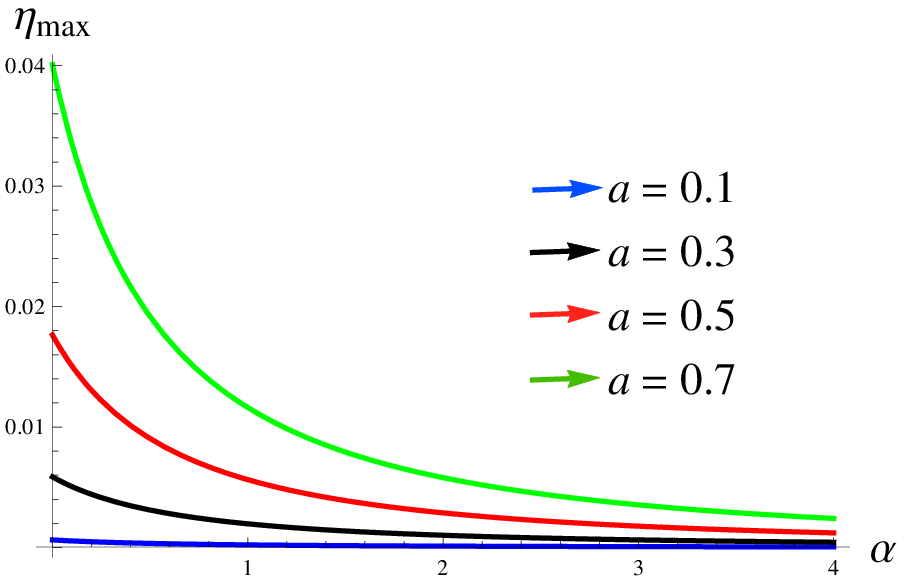}
    \end{minipage}\hspace{-0.4cm}
    \begin{minipage}[b]{0.53\textwidth}
        \includegraphics[width=0.9\textwidth]{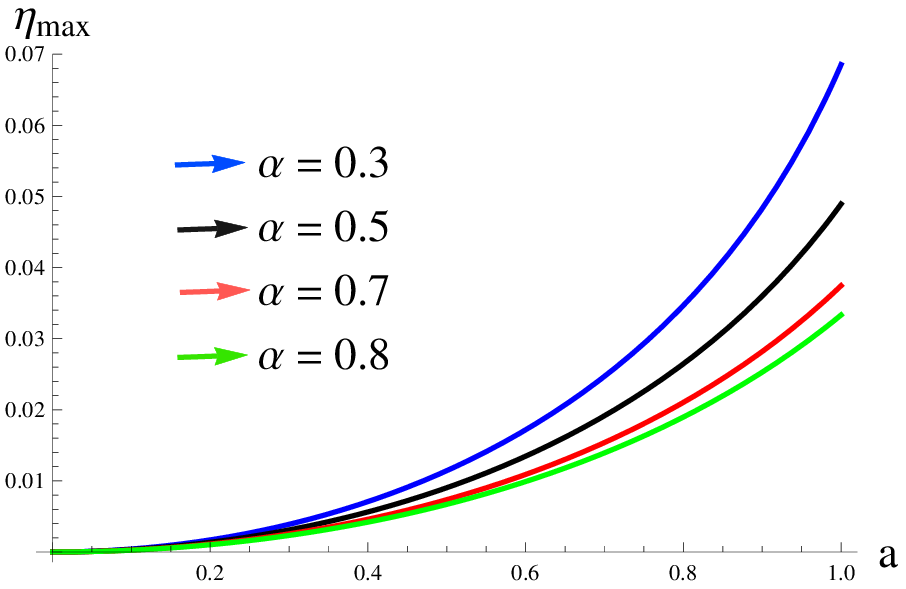}
    \end{minipage}
    \caption{Plots of maximum efficiency of energy extraction
    as a function of $a$ (left) and $\alpha$ (right).}\label{fig:9}
\end{figure*}
Consequently, Eq.(\ref{18a}) takes the form
\begin{eqnarray}\nonumber
[a^{2}&+&r^{2}-\frac{1}{r^{4}}(a^{2}(-2(1+\alpha)r+\alpha
\left(1+\alpha\right))(r^{2}+2(1+\alpha)\\\nonumber&\times&r-\alpha
\left(1+\alpha\right)))]\Omega^{2}-[\frac{1}{r^{4}}
(2a(-2\left(1+\alpha\right)r+\alpha
(1\\\nonumber&+&\alpha))(2r^{2}-2\left(1+\alpha\right)r+\alpha
\left(1+\alpha\right)))]\Omega+\frac{1}{r^{4}}
[(r^{2}-2(1\\\nonumber&+&\alpha)r +\alpha
\left(1+\alpha\right))(2r^{2}-2\left(1+\alpha\right)r+\alpha\left(1+\alpha\right))]=0.
\end{eqnarray}
Using the above equation, the angular velocity of incident particle
can be written as
\begin{eqnarray}\nonumber
\Omega_{(0)}&=&r^{4}[r^{4}(r^{2}+a^{2})-(a^{2}(-2\left(1+\alpha\right)r+\alpha
\left(1+\alpha\right))(r^{2}\\\nonumber&+&2\left(1+\alpha\right)r-\alpha
\left(1+\alpha\right)))]^{-1}\frac{1}{r^{4}}[a(-2\left(1+\alpha\right)r+\alpha\\\nonumber&\times&
\left(1+\alpha\right))(2r^{2}-2\left(1+\alpha\right)r+\alpha
\left(1+\alpha\right))]-[\frac{1}{r^{6}}(2r^{2}\\\nonumber&-&2\left(1+\alpha\right)r+\alpha
\left(1+\alpha\right))(r^{6}-2\left(1+\alpha\right)r^{5}+a^{2}
(r^{4}\\\nonumber&-&8\left(1+\alpha\right)r^{2})+\alpha
\left(1+\alpha\right)(r^{4}+8a^{2}\left(1+\alpha\right)
r-2a^{2}\\\nonumber&\times&\alpha\left(1+\alpha\right)))]^{1/2}.
\end{eqnarray}
Substituting Eqs.(\ref{22}) and (\ref{23}) into (\ref{21}), we
obtain the efficiency of energy extraction in the form
\begin{eqnarray}\nonumber
\eta=\frac{(\Omega_{(0)}-\Omega_{-})(g_{tt}+\Omega_{+}g_{t\phi})}
{(\Omega_{(+)}-\Omega_{-})(g_{tt}+\Omega_{0}g_{t\phi})}-1.
\end{eqnarray}
In order to find the maximum value of efficiency ($\eta_{max}$), the
incident particle must be splitted at the horizon of BH. Therefore,
the above equation becomes
\begin{eqnarray}\nonumber
\eta_{max}=\frac{\sqrt{1+g_{tt}}-1}{2}|_{r=r_{+}}.
\end{eqnarray}
The values of maximum efficiency of the energy extraction by Penrose
process for different values of the spin parameter $a$ as well as
parameter $\alpha$ are given in Table $\textbf{1}$. It is observed
that maximum efficiency can be enhanced as the parameter $a$
increases. The rotation of a BH has strong effects on the efficiency
of energy extraction. It is interesting to note that maximum
efficiency decreases with the increasing values of $\alpha$.
Moreover, more energy can be extracted from BH when BH rotates
rapidly. It is worthwhile to mention that energy extraction
efficiency for Kerr-MOG BH is low in comparison with the Kerr BH.
For $\alpha=0$ and $a=1$, we have the limiting value for extreme
Kerr BH, i.e., $20.7\%$ \cite{8}. So, the obtained results are the
generalization of Kerr BH. The graphical behavior of maximum
efficiency with respect to $\alpha$ (left) and $a$ (right) for
different values of $a$ and $\alpha$, respectively, is depicted in
Figure $\textbf{9}$. It is found that maximum efficiency has
decreasing behavior with the increase of $\alpha$ but increases as
spin parameter $a$ increases.
\begin{table*}
\begin{center}
\textbf{Table 1:} The maximum efficiency $\eta_{max}$ $(\%)$ of
energy extraction from BH by the Penrose process.
\begin{tabular}{c c c c c c c c c }
\hline \hline
&  {a=0.2}  &  {a=0.4}  &  {a=0.6}  &   {a=0.8}  &  {a=0.9}  &  {a=0.99}  &  {a=1.0}   \vspace{2mm}\\
\hline
 $\alpha=0.0$ & 0.255 & 1.077  & 2.705  & 5.902  & 9.009  & 16.196 & 20.711
 \\
 $\alpha=0.01$ & 0.251 & 1.060 &  2.658  & 5.776 & 8.768
  & 15.200 & 17.298&
 \\
$\alpha=0.02$  & 0.247 &1.044 & 2.612 & 5.655 & 8.538 &14.397 &15.98
&
 \\
$\alpha=0.2$  & 0.193& 0.805 & 1.970 & 4.051  &5.752 & 8.135 &8.493
&
 \\
$\alpha=0.3$  & 0.170 & 0.709 & 1.718  & 3.466  & 4.829 &6.604
&6.853 &
 \\
 $\alpha=0.4$ & 0.152 & 0.629 & 1.514  & 3.009 &4.137  &5.536 &5.724 &
 \\
$\alpha=0.5$  & 0.136 & 0.563 & 1.347 & 2.646 & 3.598 & 4.742 &4.893
&
 \\
\hline
\end{tabular}
 \end{center}
\end{table*}

\section{Concluding Remarks}

In this paper, we have investigated the neutral particle motion and
energy extraction via Penrose process around Kerr-MOG BH. The
circular motion of particles play an important role in the study of
accretion disk theory. Circular timelike geodesics are useful to
discuss the dynamics of galaxies and planetary motion. Since BHs are
basically non-emitting objects so the investigation of null
geodesics around them is also of great interest. The photons coming
from other sources carry the astrophysical information that reach to
the observer from the accretion disk. We have explored the timelike
as well as null geodesics. The effective potential approach is used
to study the stability of circular photons orbits. It is observed
that the presence of parameter $\alpha$ contributes to increase the
instability of photons orbits. The motion becomes unstable for
increasing values of $\alpha$. The rotation of BHs also effects the
stability of photons orbits. We see that the instability of orbits
increases with the increase of rotation of a BH. This situation is
much different in comparison with the mining braneworld Kerr-Newman
spacetimes where the stable circular photon orbit is almost
independent of the spin parameter $a$ \cite{15b}. We found that
circular orbits of photons around Kerr-MOG BH are more unstable in
comparison with the Kerr, Schwarzschild, and Schwarzschild-MOG BHs.

We have discussed the effective force acting on photons, which is
more attractive for larger values of spin parameter $a$. It is noted
that effective force acting on photons increases with the increasing
values of $\alpha$ and does not change as the photons move away from
the BH. We have also investigated the instability of circular
photons orbits through the Lyapunov exponent. It is observed that
Lyapunov exponent initially decreases then increases for higher
values of $a$ as compared to smaller values but does not change much
more as radial distance increases.

We have examined the energy extraction by Penrose process around
Kerr-MOG BH. It is interesting to note that negative energy
decreases with the increasing values of parameter $\alpha$. It is
worthily to mention that negative energy also decreases when a BH
rotates rapidly. We have studied the efficiency of energy extraction
from BH. We conclude that the maximum efficiency can be enhanced as
the parameter $a$ increases. The rotation of a BH has strong effects
on the efficiency of energy extraction. More energy can be extracted
from a BH when the rotation of a BH increases. Maximum efficiency
has decreasing behavior with the increase of $\alpha$. We have
compared the efficiency for Kerr-MOG BH with the Kerr BH, the
obtained results indicates that efficiency for Kerr-MOG BH is small
than that of the Kerr BH.



\end{document}